# TITLE

**Full Title**

Use of Technology and Innovations in the COVID-19 Pandemic Response in Africa

**Short Title**

COVID-19 Response in Africa

**AUTHORS & AFFILIATIONS**

Adyasha Maharana,[1] Morine Amutorine,[2] Moinina David Sengeh,[3] Elaine O. Nsoesie[4]

1. Department of Computer Science, University of North Carolina, Chapel Hill, NC, USA
2. United Nations Data Innovations Lab, Kampala, Uganda
3. Directorate of Science, Technology and Innovation, Freetown, Sierra Leone
4. Department of Global Health, School of Public Health, Boston University, Boston, MA, USA

*Corresponding author: Elaine O. Nsoesie, Department of Global Health, Boston University School of Public Health, 801 Massachusetts Ave, Crosstown Center 3rd Floor, Boston, MA 02119; onelaine@bu.edu; Tel: 617 358 3120

**ABSTRACT**

The use of technology has been ubiquitous in efforts to combat the ongoing public health crisis due to emergence and spread of the SARS-CoV-2 virus. African countries have made tremendous use of technology to disseminate information, counter the spread of COVID-19, and develop cutting-edge techniques to help with diagnosis, treatment and management of patients. The nature and outcomes of these efforts sometimes differ in Africa compared to other areas of the world due to its unique challenges and opportunities. Several countries have developed innovative technology-driven solutions to cater to a diverse population with varying access to technology. Much of the efforts are also earmarked by a flexible approach to problem solving, local tech entrepreneurship, and swift adoption of cutting-edge technology.

**INTRODUCTION**

The first cases of COVID-19 in Africa and sub-Saharan Africa were reported in Egypt and Nigeria on February 14th and 27th 2020, respectively (Figure 1).(1,2) As of September 27, 1,460,084 cases, 35,445 deaths and 1,206,237 recoveries have been reported across the continent.(3) The response to COVID-19 across Africa has varied with some countries implementing stringent physical distancing interventions and integrating data and analytics into their response, while others have not. However, since the start of the pandemic, governments, organizations and individuals have repurposed existing technologies and created novel innovations to fight COVID-19. We provide a comprehensive review of how technology and innovation has been used for enforcing physical distancing interventions, spreading awareness, tracking cases, diagnosing disease, contact tracing, clinical management of disease, and addressing misinformation.

Figure 1. Timeline of first reported cases of COVID-19 in each of the African countries.

We conducted a literature review on August 24th and 25th, 2020 using Google, Google Scholar and PubMed. We searched for articles using the following broad search terms: "[Country] AND Coronavirus AND Technology". The search was conducted for each of the 54 African countries. To supplement this search, we also searched for "Africa AND Coronavirus AND Technology" to identify articles that were missed in the country-specific searches. We included articles if they mentioned the use of technology in the control of COVID-19 in any of countries. Due to the novelty of the pandemic, most of the relevant literature identified were not found in peer reviewed publications. The included articles were from trusted news sites, government, and organization websites such as, the United Nations and the World Health Organization. The articles included in this review can be grouped into (1) disease prevention, (2) tracking cases, (3) healthcare and (4) addressing misinformation.

**MAIN TEXT**

**Disease prevention**

*Enforcing Physical Distancing Interventions*

Some African countries implemented physical distancing interventions before and after confirming the first COVID-19 case. The interventions ranged from border closures to lockdowns lasting days, weeks or months.(4) Enforcing these physical distancing interventions has been challenging and countries have adopted technology in different ways to "flatten the curve". For example, Ethiopia introduced an internet-based transportation payment system to enforce distancing for bus travelers.(5) Similarly, the government of Sierra Leone developed an electronic pass (e-pass) management system to limit movement to essential workers (i.e., individuals needed to maintain essential services such as, supply of food, medical resources, gasoline, electricity, water and telecommunications) during the national lockdown. Individuals wishing to move could apply for an e-pass and receive confirmation of acceptance or rejection via SMS within 12 hours.(6) The Sierra Leone government also used drones to capture images to assess citizen compliance during lockdown in Freetown,.(7) Drones were also used to enforce lockdowns and physical distancing in South Africa, Morocco and Tunisia by issuing warnings and announcements, and questioning people on the street about their reasons for being out.(8–10)

*Awareness and Hygiene*

From the start of the pandemic, African countries have used innovative approaches to spread awareness and hygiene messages. For example, the Rwandan National Police deployed drones to spread COVID-19 awareness messages, while in Ghana, drones were used for disinfecting public places.(11,12)

Startups, governments and UN organizations also used existing technologies to support COVID-19 response efforts. Existing tech-enabled supply chains (e-commerce) were used to source and distribute hygiene equipment to various parts of Kenya.(13) Ushahidi, a decade-old Kenyan tech innovation, was used worldwide to ask and receive help during lockdown, gather information about the virus and report contamination.(14) A startup in Morocco released a bot which answered questions regarding COVID-19 in Arabic.(15) In Uganda, the UN Innovations Data Lab used machine learning to deploy a text-to-speech radio monitoring technology to gather first-hand incidents and local opinions and beliefs about the pandemic. The data gathered could be used to identify areas with greatest need of healthcare resources, combat misinformation and gauge the social and economic impacts of lockdown.(16) While in Guinea, the government capitalized on bulk SMS messaging and broadcasted information through caller ringtones in the various regional languages.(17)

There were also locally designed interventions used to spread awareness and hygiene messaging. For example, an automatic hand sanitizer dispenser developed in Senegal was used to reduce the need for supervised handwashing.(18) A contactless electrical soap dispenser was developed and used in communities in Ethiopia.(19) Similarly, a solar-powered hand washing sink was developed by local innovators in Ghana and a touchless handwashing station was developed and distributed in Uganda.(20,21)

*Medical Supplies*

To address the challenge posed by a lack of medical supplies, some countries have adopted technology in the printing and distribution of certain medical supplies such as, masks, test kits and ventilators. 3D printing and drones have been used to make and deliver medical supplies, respectively. For instance, drones were used to deliver medical products and testing samples to labs in Rwanda and Ghana, respectively.(22) Drones were also used to deliver personal protective equipment (PPE) and COVID-19 test kits in these countries.(23,24)

In Kenya, the 3D printing community printed affordable PPE, plastic face shields and ventilator adaptors that allow doctors to treat two to four patients at the same time.(25) 3D printing was used to develop personal protective equipment in Cameroon, Egypt and the Gambia.(26–28) Local inventors in Uganda created plastic motorcycle shields to keep motorcycle taxis running during the pandemic.(29)

There have also been efforts to locally develop ventilators. For example, a cheap, portable, automatic ventilator made by a Nigerian student was used in COVID-19 wards to combat the shortage of ventilators.(30) Similarly, a young inventor in Ethiopia created a ventilator that used a plastic pouch, mechanical ventilator and a screen operated from a phone.(31) Students at the Fourah Bay College in Sierra Leone also developed a mechanical ventilator.(32)

Furthermore, in response to the COVID-19 pandemic, the African Union created the Africa Medical Supplies Platform, a collaborative platform that operates across national boundaries and partners with African governments at all levels to connect COVID-19 response teams with medical supplies.(33,34)

**Tracking cases**

*Assessing Risk and Tracking Cases/Burden*

Several countries developed dashboards that were used for monitoring the number of cases and deaths.(35–38) In Rwanda, a GIS system was used to track cases at the household level and the data was used to assess the need for lockdown measures and other public-health

interventions.(39) Post lockdown, South Africa adapted a technology used for identifying rhinoceros poaching hotspots and integrated with health and location data to spot new COVID-19 case clusters in their community.(40)

Sierra Leone launched a symptom checker and COVID-19 information resource that can be used with and without the internet.(41) The system can be accessed via SMS and a previously launched USSD (Unstructured Supplementary Service Data) portal which was developed to provide timely information to its residents. On April 26, there had been about 250,000 USSD responses.(42) Residents in Ghana could dial a USSD code on a phone after which, they were prompted to answer questions about their symptoms, age, travel history etc. The data collected was aggregated and analyzed to predict if the person was likely to get the virus. This tool was accessible by most people since it did not require a smartphone to operate an app or website.(43) Additionally, a COVID-tracking app was also released in Ghana to track cases based on location, enforce adherence to quarantine rules and reach out to people who needed to quarantine.(44) Humanitarian agencies also used satellite imagery and data to map and identify high-risk areas to generate situational awareness and capacity.(26) Researchers in Burkina Faso used phenomenological models to characterize the growth of the COVID-19 outbreak and forecast future growth. (45)

Tunisian agencies used Tableau dashboards to analyze calls made to the country's general emergency number (which saw two years' worth of calls in less than a month due to the pandemic) in order to find patients who needed immediate attention.(46) In Nigeria, a digital triage tool was used to assess the user's risk to COVID-19 based on their answer to a curated

questionnaire.(47) A WhatsApp bot developed in South Africa was used to deliver health alerts to people.(48) In Algeria, a free mobile self-diagnosis and case-monitoring application was developed and deployed in Arabic.(49) Ethiopia has developed dedicated mobile phone apps for recording personal identification information at ports of entry and followed it up with the traveler's temperature. This information was further used by a COVID-19 surveillance and tracking system for contact tracing activities. Additionally, a data collection tool was developed to gather data during door-to-door COVID-19 screening campaigns.(50)

*Contact Tracing*

The official triaging app in Ghana was used to augment the contact tracing efforts of the government by recording the location of app users.(51) In South Africa, handsets equipped with location-tracking technology were distributed in the hardest-hit areas to perform contact tracing.(52) Tunisia launched a contact tracing mobile phone app which alerted users if they have been in close contact with someone who tested positive for COVID-19.(53) In Rwanda, contact tracing was performed by tracing infections through the paperless Open Data Kit application.(54)

Genomic epidemiology has also played a significant role in understanding the spread of COVID-19 in some countries. Ghanaian scientists were among the first to sequence the SARS-Cov-2 in Africa using local resources.(55) Zambian scientists used nanopore sequencing to track mutations in SARS-CoV-2, understand the origin of cases and track people in the chain of infection.(56) Similarly, Gambian scientists traced the origin of the first six cases in their country by sequencing the virus strain from cases and mapping it to eleven other strains of Asian, American and European origin.(57)

*Diagnosing Disease*

Testing capacity remains a challenge in many African countries. A lab in Senegal has developed at-home test kits for COVID-19 that could cost as low as $1 per kit and can provide results in ten minutes.(58) They are undergoing trials and can be rapidly produced in massive amounts, if given the go-ahead. Furthermore, sample pooling for COVID-19 testing which involves testing samples from many people at once has been used in some countries. The growing pressure to test more people for COVID-19 has made pooling an attractive option since it conserves chemical reagents, time and money. Researchers in Rwanda have come up with an improved manner of group testing that could reduce the cost of testing from 9 USD to 0.75 USD per person.(59) In response to Ebola, Measles and now COVID-19 outbreaks, Democratic Republic of Congo (DRC) has been steadily ramping up investment in laboratory capacity and more importantly, genomic surveillance with DNA databases in order to trace the path of transmission and explain dynamics of the outbreak.(60) Furthermore, an open-source software developed in Tunisia was used to compare chest X-rays of patients suspected of having the virus to X-rays of patients with confirmed COVID-19 disease.(61)

Home-grown swab tests in Uganda have brought down the prohibitive 65USD per test cost of COVID-19 testing. These tests can also be used for point-of-care deployment in remote areas.(62) Makerere University in Uganda has built innovations like thermal imaging for detecting COVID-19.(63)

**Healthcare**

*Clinical Management*

Telehealth has been adopted in several countries for clinical management. For example, telehealth has gained traction in Uganda to prevent congestion and spread of COVID-19 in healthcare facilities.(64) In Eswatini, ongoing in-person care of TB and HIV patients was adapted for no-contact care by asking patients to film a video while taking the required medications and send it to a nurse who monitors their health.(65) In Rwanda, a digital-first solution was developed to encourage COVID-19 patients to self-care at home instead of overcrowding their healthcare system, leaving the urgent care free for critically ill patients.(66) A telehealth platform, created by a Libyan entrepreneur to connect healthcare professionals in the diaspora to patients in their home countries, became the go-to app for triaging of COVID-19 patients in Libya.(67) Telehealth usage surged in Ghana and demonstrated that time-taking, manual clinical management can be replaced with a less stressful technology-driven solution.(68) Egypt's broadband network has been leveraged to deploy telehealth solutions for primary care units.(69)

Several innovative tools were used throughout Africa in the clinical management of patients. For example, in a bid to minimize contact of health workers with COVID-19 patients, robots were deployed in healthcare settings in Rwanda to perform simple tasks like taking temperature, monitoring the health of patients, delivering food and medicine etc.(70) Similarly, a hospital in Tunisia used a robot to take the pulse, check temperature and blood oxygen levels of patients to limit contact between patients and clinicians.(71,72) Student engineers in Senegal used their technical skills to create inventions to alleviate the burden of the COVID-19 problem. One such example was in the designing of a small robot called 'Dr. Car' that would measure the temperature and blood pressure of infected patients at hospitals, reducing the risk of exposure to healthcare workers. Additionally, the robot could also ease patient-doctor communications during treatment, especially in hard to reach rural areas.(73)

Digital solutions have played an important role in the care and management of health for COVID-19 as well as non-COVID patients. For example, an app for pregnant mothers in Uganda provided up-to-date guidelines on practicing social distancing during pregnancy, handling unexpected situations and healthcare information.(74) The Moroccan Ministry of Health introduced an app platform for healthcare professionals to promote fast exchange of knowledge on COVID-19 with other healthcare professionals.(75) A healthcare platform in Benin was used to connect hospitals and different healthcare stakeholders on a single platform saving the time spent on procuring patients' medical records.(76) A voice and SMS messaging system for HIV patients was repurposed by Ugandan researchers to support quarantined individuals and flag potential concerns.(77)

Other innovations include, medicine delivery by connecting COVID-19 patients with pharmacies in Egypt and ride-hailing apps that offered doorstep delivery services for medical necessities such as contraceptives, pregnancy and HIV tests in Uganda.(78,79)

**Addressing misinformation**

COVID-19 misinformation has been considered a major issue affecting the control of the pandemic. Inaccurate information on the SARS-CoV-2 virus and COVID-19 disease has prevented people from following the correct preventive guidelines and prompted them to accept hoax cures. For instance, in Namibia, elephant dung was briefly sold for exorbitant prices with a claim to cure COVID-19.(80)

To address misinformation, several African governments, organizations and communities have launched efforts that use technology and context-specific approaches.(81) The Nigerian Health Ministry used WhatsApp push notifications to broadcast accurate information about COVID-19, its treatment, symptoms and prevention.(82) Residents in Sierra Leone could verify information seen on social media and other sources through their USSD portal, which can assist in combating misinformation.(83) Relatedly, Benin set up an official digital platform which was accessible to all its citizens, to broadcast accurate information on COVID-19. It has a flash news functionality to respond to instantaneous fake news.(84) The Moroccan government passed a controversial bill to stem the spread of fake news by making it illegal and detainable by law.(85)

There were also opportunities for youths to be involved in combating misinformation. For example, a knowledge synthesis team in Ethiopia curated and analyzed information on COVID-19 daily and produced updated treatment and prevention guidelines which were then published by the Ethiopian Ministry through all available social media channels in order to combat misinformation.(86) Similarly, Zimbabwean youths combed through online comments on social media in order to dispel incorrect statements made by the general public about COVID-19.(87)

Volunteers in Sudan used online platforms to educate the public about COVID-19 and dispel rumors.(88) Niger also launched an audio-visual campaign and trained numerous young people to make videos that counter fake news or misinformation on COVID-19 and disseminate them online.(89)

Radio and other media (including, TV and social media) have been used in Zambia, Mali and Uganda to debunk fake news, and deliver reliable and accurate information on the pandemic in local languages.(90–92) In Ghana, young healthcare workers and volunteers visit high risk areas to spread information.(92) When technology was not an option, such as for some areas in Chad where families did not have resources to own a radio, traditional storytellers (troubadours) have been spreading COVID-19 prevention messages to combat rumors. Awareness posters were also used to serve such off-the-grid areas.(93) While most efforts on fighting misinformation have been commendable, there has also been troubling news that laws passed by some African countries in order to curb misinformation were used to curb citizen privacy and press freedom.(94)

**Challenges**

While technology has been adopted to address many COVID-19 related concerns, it is difficult to measure their impact on controlling the pandemic. Furthermore, challenges still persist that limit the adoption of new technologies. For example, tele-health has gained traction as a safe alternative to in-person clinical visits; however, its widespread adoption is impeded by internet connectivity (or lack thereof) and geographical limitations to accessing medicine deliveries and lab sample pick-ups.(95) The COVID-19 pandemic has brought a focus to the need for data, digital connectivity and technology among countries in Africa. For digital health solutions to succeed and be accessible to everyone, barriers to widespread distribution of the internet needs to be addressed and all stakeholders must cooperate to come up with a resilient, affordable, scalable, long-term plan that enables rapid deployment.

Furthermore, local innovations continue to highlight the need to invest in and adopt "home grown" technology that can assist in the control of future outbreaks and improve public health across the continent. Solutions to health crises should not be left to local and international experts, youth and community health workers should be included in pandemic planning and response conversations. The median age across the African continent is 19 years, and many of these young people have innovative ideas and are passionate about developing technology and using data to address local problems.(96) For example, tech startups are being funded in Nigeria to come up with innovative solutions that promote primary care, improve efficiency of health processes, communicate health risks, assist patients with self-triage etc.(97) Morocco has a dedicated industrial base for unmanned flight technology, however most of its drones are imported from China. The pandemic has prompted the rise of startups with the aim to build drones in Morocco for thermal surveillance and disinfectant-spraying.(98) Several startups in Morocco have been funded to assist with solutions for tackling COVID-19, including one that uses 3D printers to develop masks, PPE and ventilators for hospitals. The Tunisian entrepreneurial ecosystem financed several scientific ideas which have promising impact on the pandemic crisis.(99) Egyptian IT Agency's dedicated innovation and entrepreneurship center supported hackathons for building different parts of a COVID-19 response system that helps people, disrupted businesses and health agencies.(100) If sustained, these technological advances can lead to long-term improvements in public health infrastructure in Africa.

Furthermore, limitations identified during this pandemic can be used to spur the development of better public health surveillance systems. It has been recommended to create genomic hubs in well-placed locations throughout the continent so that Africa can rapidly generate and curate genomic data related to outbreaks.(101) Point-of-care diagnostic services (on-location healthcare including testing, scanning etc.) supplemented with machine learning and information systems should be prioritized in combating the ongoing pandemic and future outbreaks, since they have proved their necessity in previous healthcare epidemics like HIV.(102) The development of reliable disease detection systems including, diagnostics, monitoring and developing well-equipped health facilities can help countries to prepare for future outbreaks.(103)

**CONCLUSION**

Africa has a long history of combating disease epidemics, which has proven extremely useful in combating the COVID-19 pandemic across the continent. This experience in fighting epidemics has been channeled to build resilient public health responses to COVID-19 elsewhere in the world.(104) However, to create significant advances in healthcare, African nations must invest in technologies (such as AI, satellite imagery, drones, robotics, 3D printing, cloud computing). This should include investment in human capital, and establishing legal and regulatory frameworks that govern the use of these technologies. This will facilitate innovation, encourage financial investment and foster entrepreneurship. Combining technology with well-established public health techniques for combating epidemics would enable governments to solve problems quicker and more robustly if used sensibly and ethically.

8. Coronavirus: Tunisia deploys police robot on lockdown patrol - BBC News [Internet]. [cited 2020 Oct 10]. Available from: https://www.bbc.com/news/world-africa-52148639

9. Morocco launches fleet of drones to tackle virus from the sky | AFP - YouTube [Internet]. [cited 2020 Oct 19]. Available from: https://www.youtube.com/watch?v=Sz5bOnEP8Vk&ab_channel=AFPNewsAgency

10. WATCH | Limpopo mayor becomes Big Brother, thanks to a drone [Internet]. [cited 2020 Oct 19]. Available from: https://www.sowetanlive.co.za/news/south-africa/2020-04-12-watch--limpopo-mayor-becomes-big-brother-thanks-to-a-drone/

11. Sarfo AK, Karuppannan S. Application of Geospatial Technologies in the COVID-19 Fight of Ghana. Transactions of the Indian National Academy of Engineering [Internet]. 2020 Jun 4 [cited 2020 Oct 10];5(2):193–204. Available from: https://doi.org/10.1007/s41403-020-00145-3

12. COVID-19 response in Rwanda: Use of Drones in Community awareness | WHO | Regional Office for Africa [Internet]. [cited 2020 Oct 19]. Available from: https://www.afro.who.int/news/covid-19-response-rwanda-use-drones-community-awareness

13. Home - Safe Hands Kenya [Internet]. [cited 2020 Oct 4]. Available from: https://www.safehandskenya.com/

14. Ushahidi in the era of COVID 19 - Ushahidi [Internet]. [cited 2020 Oct 10]. Available from: https://www.ushahidi.com/blog/2020/03/30/ushahidi-in-the-era-of-covid-19

15. Digitalisation and the coronavirus in Morocco: From care to control? | Heinrich Böll Stiftung | Brussels office - European Union [Internet]. [cited 2020 Oct 10]. Available from: https://eu.boell.org/en/2020/04/09/digitalisation-and-coronavirus-morocco-care-control

16. Using speech-to-text technology to support response to the COVID-19 pandemic • UN Global Pulse [Internet]. [cited 2020 Oct 10]. Available from:
17

https://www.unglobalpulse.org/2020/05/using-speech-to-text-technology-to-support-response-to-the-covid-19-pandemic/

17. Innovative tech and connectivity key to fighting COVID-19 in Africa | Africa Renewal [Internet]. [cited 2020 Oct 10]. Available from: https://www.un.org/africarenewal/news/coronavirus/innovative-tech-and-connectivity-key-fighting-covid-19-africa

18. Senegal's engineering students design machines to fight Covid-19 [Internet]. [cited 2020 Oct 10]. Available from: https://www.france24.com/en/20200513-senegal-s-engineering-students-design-machines-to-fight-covid-19

19. Young inventor helps Ethiopia's COVID-19 crisis | World| Breaking news and perspectives from around the globe | DW | 05.05.2020 [Internet]. [cited 2020 Oct 10]. Available from: https://www.dw.com/en/young-inventor-helps-ethiopias-covid-19-crisis/a-53334966

20. Coronavirus: Ten African innovations to help tackle Covid-19 - BBC News [Internet]. [cited 2020 Oct 10]. Available from: https://www.bbc.com/news/world-africa-53776027

21. Mak unveils a Touchless Handwashing Kit for public shared spaces in response to COVID-19 pandemic - Makerere University News [Internet]. [cited 2020 Oct 10]. Available from: https://news.mak.ac.ug/2020/08/mak-unveils-a-touchless-handwashing-kit-for-public-shared-spaces-in-response-to-covid-19-pandemic/

22. Five ways humanitarians use technological innovation to deliver during COVID-19 | by United Nations OCHA | Humanitarian Dispatches | Medium [Internet]. [cited 2020 Oct 10]. Available from: https://medium.com/humanitarian-dispatches/five-ways-humanitarians-use-technological-innovation-to-deliver-during-covid-19-40ce8e977fc4

23. Medical delivery drones are helping fight COVID-19 in Africa, and soon the US | World Economic Forum [Internet]. [cited 2020 Oct 10]. Available from: https://www.weforum.org/agenda/2020/05/medical-delivery-drones-coronavirus-africa-us
18

64. Kamulegeya LH, Bwanika JM, Musinguzi D, Bakibinga P. Continuity of health service delivery during the COVID-19 pandemic: the role of digital health technologies in Uganda. Pan African Medical Journal. 2020 May 20;35(Supp 2).

65. Responding to coronavirus COVID-19 in Eswatini | MSF [Internet]. [cited 2020 Oct 4]. Available from: https://www.msf.org/responding-covid-19-eswatini

66. Coronavirus and Africa - in Rwanda, Technology to the Rescue | Institut Montaigne [Internet]. [cited 2020 Oct 4]. Available from: https://www.institutmontaigne.org/en/blog/coronavirus-and-africa-rwanda-technology-rescue

67. This Doctor's App Is Helping Libya Triage Its Coronavirus Patients [Internet]. [cited 2020 Oct 4]. Available from: https://www.forbes.com/sites/andrewwight/2020/04/02/how-did-a-libyan-doctor-give-back-a-covid-19-fighting-app/#6f75ee0b235d

68. COVID-19 Drives Health Care Tech Innovation in Ghana | Voice of America - English [Internet]. [cited 2020 Oct 4]. Available from: https://www.voanews.com/covid-19-pandemic/covid-19-drives-health-care-tech-innovation-ghana

69. Innovative tech and connectivity key to fighting COVID-19 in Africa | Africa Renewal [Internet]. [cited 2020 Oct 4]. Available from: https://www.un.org/africarenewal/news/coronavirus/innovative-tech-and-connectivity-key-fighting-covid-19-africa

70. Rwanda has enlisted anti-epidemic robots in its fight against coronavirus - CNN [Internet]. [cited 2020 Oct 4]. Available from: https://www.cnn.com/2020/05/25/africa/rwanda-coronavirus-robots/index.html

71. Tunisia: Intelligent Robot Jasmin to Help Fight COVID-19 in Tunisia - allAfrica.com [Internet]. [cited 2020 Oct 11]. Available from: https://allafrica.com/stories/202007110148.html
24

102. Mashamba-Thompson TP, Drain PK. Point-of-Care Diagnostic Services as an Integral Part of Health Services during the Novel Coronavirus 2019 Era. Diagnostics [Internet]. 2020 Jul 3 [cited 2020 Oct 1];10(7):449. Available from: https://www.mdpi.com/2075-4418/10/7/449

103. How Senegal is confronting the challenge of COVID-19 [Internet]. [cited 2020 Oct 1]. Available from: https://theconversation.com/how-senegal-is-confronting-the-challenge-of-covid-19-134128

104. Using Reverse Innovation to Fight Covid-19 [Internet]. [cited 2020 Oct 1]. Available from: https://hbr.org/2020/06/using-reverse-innovation-to-fight-covid-19